# Enhanced spin-orbit torque via modulation of spin current absorption

Xuepeng Qiu[1,2†], William Legrand[1†], Pan He[1], Yang Wu[1], Jiawei Yu[1], Rajagopalan Ramaswamy[1], Aurelien Manchon[3§], and Hyunsoo Yang[1]*

[1]*Department of Electrical and Computer Engineering, and NUSNNI, National University of Singapore, 117576, Singapore*

[2]*Shanghai Key Laboratory of Special Artificial Microstructure Materials & School of Physics Science and Engineering, Tongji University, Shanghai 200092, China*

[3]*Division of Physical Science and Engineering, King Abdullah University of Science and Technology (KAUST), Thuwal 23955, Saudi Arabia*

[†]These authors contributed equally to this work.
*eleyang@nus.edu.sg,
[§]aurelien.manchon@kaust.edu.sa

The magnitude of spin-orbit torque (SOT), exerted to a ferromagnet (FM) from an adjacent heavy metal (HM), strongly depends on the amount of spin currents absorbed in the FM. We exploit the large spin absorption at the Ru interface to manipulate the SOTs in HM/FM/Ru multilayers. While the FM thickness is smaller than its spin dephasing length of 1.2 nm, the top Ru layer largely boosts the absorption of spin currents into the FM layer and substantially enhances the strength of SOT acting on the FM. Spin-pumping experiments induced by ferromagnetic resonance support our conclusions that the observed increase in the SOT efficiency can be attributed to an enhancement of the spin-current absorption. A theoretical model that considers both *reflected* and *transmitted* mixing conductances at the two interfaces of FM is developed to explain the results.

The manipulation of the magnetization by electrical currents is one of the ultimate goals in spintronics [1-4]. Conventional spin transfer torque (STT) utilizes the spin polarized current to transfer angular momentum from one FM to another FM [5,6], and it has been applied to various spintronic devices such as a magnetic random access memory (MRAM) [7-9] and race-track domain wall memory [10]. Recently emerging SOT opens a new prospect in the magnetization manipulation of the HM/FM bilayers [1,11-25]. In a SOT device, the electrons flowing in a HM layer with strong spin-orbit coupling are spin polarized and the subsequent spin current transfers its angular momentum to the adjacent FM.

Building energy efficient SOT devices essentially relies on the magnitude of SOT. Prior efforts to achieve enhanced SOT are mostly based on the engineering of spin orbit coupling strength [11-13,15-17,19,26,27], through either the bulk spin Hall or interfacial Rashba effect. While these previous works focused on generating a stronger spin current, very little attention has been paid to modulate the spin current absorption to manipulate torques exerted on the FM layer. A few recent works have been devoted to study the role of transparency in the HM/FM interface on the amount of spin currents diffusing into the FM [14,28]. By varying the FM material adjacent to HM, the interface transparency changes, which in turn influences the total torque exerted on the FM layer. However, the absorption of spin currents after they enter the FM layer has yet to be considered in the framework of SOT manipulation, both experimentally and theoretically. It has been shown that the spin current absorption inside a FM saturates beyond a characteristic length due to the spin scattering events [29]. Thus, modifying the level of spin scattering is expected to greatly modulate the absorption of spin currents inside an ultrathin FM and provides a new prospect for SOT manipulation.

In this work, we show a large enhancement of SOT utilizing a thin Ru capping layer on top of Pt/FM. Instead of relying on engineering of spin orbit coupling, we demonstrate a control of the overall spin torque magnitude in the ultrathin FM by exploiting the spin current absorption at the FM/Ru interface. Traditionally, Ru has been a central material for a synthetic antiferromagnet (SAF) structure, which has been widely used in magnetic sensors and recording media due to its high thermal stability, small net magnetic moment, and negligible fringing magnetic fields [30-32]. An enhancement of giant magnetoresistance utilizing a SAF (CoFe/Ru/CoFe) free layer and a substantial reduction of STT switching currents by incorporating a thin Ru layer in magnetic nanopillars have been reported, as Ru is a strong spin scatterer and thus increases the spin polarization inside the spin valves [33,34]. Recently, a very high domain wall speed has been realized in Pt/SAF structures [35], in which the effect has been mainly attributed to the exchange coupling torque between two FMs across a thin Ru layer. While Ru has been one of the most important materials in spintronics, our demonstration of a large enhancement of SOT with Ru can find an immediate implementation in constructing advanced spintronic devices.

All the films are deposited by ultra-high vacuum sputtering with a base pressure < $2\times10^{-9}$ Torr. Four series of samples are deposited. The first series of samples are for the study of Ru thickness dependence with the structure of Si substrate/2 MgO/4 Pt/0.4 Co/0.3 Ni/0.1 Co/$t_{Ru}$ Ru/1.2 MgO/3 SiO$_2$ (thicknesses are in nm), where $t_{Ru}$ is varied from 0 to 2 nm. The second series of samples are for the study of FM thickness dependence with the structure of Si substrate/2 MgO/4 Pt/$t_{Co}$ Co/0.4 Ni/0.1 Co/0.6 Ru/1.2 MgO/3 SiO$_2$, where $t_{Co}$ is varied from 0.4 to 0.9 nm, so the total FM thickness $t_{FM}$ is varied from 0.9 to 1.4 nm. The third series of samples are for the study of FM thickness dependence with the structure of Si substrate/2 MgO/4 Pt/$t_{Co}$ Co/0.4 Ni/0.1 Co/1.2 MgO/3 SiO$_2$, where $t_{Co}$ is varied from 0.4 to 0.9 nm, so the total FM thickness $t_{FM}$ is varied from 0.9 to 1.4 nm. The fourth series of samples are for ferromagnetic resonance (FMR) measurements with the structure of Si substrate/3 Ta/2 Cu/[0.3 Co/0.6 Ni]$_3$/0.1 Co/2.4 Cu/0.4 Co/0.3 Ni/0.1 Co/$t_{Ru}$ Ru/1.2 MgO/3 SiO$_2$, where $t_{Ru}$ is varied from 0 to 1.4 nm. A reference sample for FMR is also deposited with the structure of Si substrate/3 Ta/2 Cu/[0.3 Co/0.6 Ni]$_3$/0.1 Co/2.4 Cu/1.2 MgO/3 SiO$_2$. All the samples possess perpendicular magnetic anisotropy (PMA), whose magnetization predominantly orients perpendicular to the film plane [36]. Except for FMR, the films are subsequently patterned into the devices with a 10 μm width in both the channel and Hall bar by photolithography and Ar ion-milling. All the measurements are performed at room temperature.

In order to study the effect of a Ru capping layer on SOT, we first measure the devices with different thicknesses of the Ru capping layer on top of Pt/Co/Ni/Co multilayers. A schematic of the film structure is shown in Fig. 1(a). Figure 1(b) shows current induced SOT switching in a series of samples with different Ru thicknesses. A pulsed dc current of a duration of 200 μs is applied to the devices and the Hall voltage is measured after a 100 μs delay. The pulsed dc current with an interval of 0.1 s is used to eliminate the accumulated Joule heating effect. For simplicity, the current is assumed to flow uniformly throughout all the metallic layers. A fixed 1 kOe magnetic field is applied along the positive current direction. For increasing the Ru thickness ($t_{Ru}$), the switching current first decreases to a minimum value at $t_{Ru}$ = 0.6 nm and then increases. Figures 1(c-

e) summarize the switching current density ($J_S$), SOT switching efficiency $\eta = (H_{an} - H_{ext})/J_S$, and SOT effective fields, respectively, for different Ru thicknesses. $H_{an}$ is the perpendicular anisotropy field and $H_{ext}$ is the external magnetic field along the current direction required to assist the deterministic magnetization switching. The longitudinal ($H_L$) and transverse ($H_T$) SOT effective fields are obtained from the harmonic Hall voltage measurements [17,22,37,38]. The SOT switching efficiency and SOT effective fields are maximized at $t_{Ru}$ = 0.6 nm, where the switching current density acquires a minimum value. By introducing a 0.6 nm thick Ru layer on top Pt/Co/Ni/Co structure, a 1.3 times enhancement in $H_L$ and 1.7 times enhancement in $H_T$ are achieved (Fig. 1(e)) which leads to a substantially improved SOT switching efficiency (Fig. 1(d)).

The observed large enhancement of SOT induced by Ru can in principle originate from the Pt/FM interface, the bulk of FM, or the FM/Ru interface. SOT generated by 3$d$ FM itself can be first excluded as it is known to be negligible [39]. The increase in the SOT switching efficiency cannot be attributed to the changes in $H_{an}$ for different $t_{Ru}$, as the minimum value of $H_{an}$ is obtained at $t_{Ru}$ = 0.2 nm [36]. As the magnetic reversal occurs via domain nucleation/expansion in the sample [25,40,41], the pinning field for different $t_{Ru}$ is also examined [36]. The pinning field shows almost no change for 0.2 nm < $t_{Ru}$ < 1 nm and thus can be excluded as the reason for the observed large enhancement of SOT. The saturation magnetization ($M_S$) measurements show that the density of states of the FM/Ru interface is modified [42,43], resulting in a reduction of $M_S$. In order to examine whether the reduction of $M_S$ can explain our data, we replace Ru by Rh [36]. In this case, $M_S$ is also modified, but no correlation is found with the torque, indicating that for both Rh and Ru capping, the reduction in $M_S$ cannot be the main origin of the enhancement of the torque.

Moreover, the spin Hall angle of Ru is measured to be negligibly small and has the same sign (positive) compared to Pt [36], counteracting the SOT from Pt. In contrast, we observe a large enhancement of SOT due to the Ru layer, which cannot be accounted for by the above scenarios.

A remaining plausible origin is that the spin current absorption in the FM is enhanced by the additional Ru layer. We have evaluated this possibility by changing the thickness of the FM. Figures 2(a,b) show the current induced SOT switching measurements for Pt/Co/Ni/Co/Ru (Ru capped) and Pt/Co/Ni/Co/MgO (MgO capped) heterostructures, respectively, with varying the FM thickness ($t_{FM}$). One can observe that for small FM thicknesses, the switching behavior and associated SOT effective fields largely differ between Ru and MgO capped samples, but approach a similar trend on increasing the FM thickness. Figures 2(c-e) compare the SOT switching efficiency and SOT effective fields versus $t_{FM}$ in Ru and MgO capped samples. The SOT efficiency and effective fields for both series converge to similar values at $t_{FM} \sim 1.2$ nm, above which the Ru capping induced SOT enhancement is no longer effective. Recent magnetotransport measurements have shown that the absorption of transverse spin currents is proportional to the FM thickness with a characteristic saturation length of 1.2 nm [29]. For FM thicknesses larger than 1.2 nm, the spin current is fully absorbed in the FM, therefore there is no enhancement of the SOT.

Figure 3 illustrates the proposed mechanism of Ru induced SOT modulation by comparing the two different cases for MgO and Ru capping. In the case of Pt/FM/MgO, the transverse spin currents from the Pt layer, which are not fully absorbed in FM, are reflected back at the FM/MgO interface. However, in the Pt/FM/Ru structure, Ru acquires negative spin polarization at the interface [42,43], which could enhance transverse spin current absorption. Indeed, it is well known from spin relaxation studies that the rapid change in magnetic texture is a source of spin dephasing [44]. The necessity of a Ru spin scatterer to induce the SOT enhancement is elucidated by replacing Ru with Cu. In this latter case, no SOT modulation is observed [36]. In fact, both Co/Ru and Co/Cu interfaces induce longitudinal spin relaxation, a phenomenon known as spin memory loss (SML) [45,46]. Although the magnitude of SML at Co/Ru and Co/Cu interfaces is similar (the phenomenological δ parameter accounting for such interfacial spin relaxation is measured to be ~0.3 for both [47]), the effects of the Ru and Cu capping layers differ completely in our experiments. This can be due to the fact that the interfacial SML parameter extracted from the experiments [45-47] applies on the spin component that is longitudinal to the magnetization direction. In contrast, our experiment addresses spin transfer torque, which concerns the transverse spin component. As the magnetic texture at FM/Ru interface changes abruptly [42,43], it is possible that the transverse spin absorption is substantially enhanced, resulting in an increased torque.

In order to further investigate the modulation of the spin current absorption by the Ru layer, spin-pumping experiments induced by FMR are performed by a field-modulated technique [29,36]. The measurement setup is shown in Fig. 4(a). In the structure of FM1/Cu/FM2/Ru ($t_{Ru}$), the bottom Pt layer is replaced by a thick perpendicularly magnetized Co-Ni multilayer (FM1), which allows the FMR detection of the spin current. In such a spin-valve design, the precession of the magnetization inside the thick Co-Ni layer pumps a spin-current into the thin 0.8 nm FM2 through the Cu spacer. The more the spin current is absorbed in the thin FM2 layer, the more spins are pumped from the FM1 layer, leading to a broadening of the FM1 resonance peaks.

For example, Fig. 4(b) shows two resonance peaks with $t_{Ru}$ = 0 and 0.6 nm. The linewidth of the resonance in the Ru capped sample increases, which is an indication of enhanced damping in the FM1 due to more pumped spin currents into FM2/Ru. Figure 4(c) shows the frequency dependence of the full width at half-maximum (FWHM) of the resonance with different Ru thicknesses ($t_{Ru}$ = 0, 0.6, and 1.4 nm), and the damping parameter (α) in the thick FM1 is extracted. The inset in Fig. 4(c) shows the damping parameter ($α_{ref}$) for the reference sample without the FM2. The difference in the damping parameter with the reference sample, $Δα = α − α_{ref}$, which is proportional to the spin current absorption in FM2, is shown as a function of $t_{Ru}$ in Fig. 4(d). The peak in Δα at $t_{Ru}$ = 0.6 nm corresponds to the thickness for which the maximal efficiency of SOT in the Pt/FM/Ru heterostructure is observed in Fig. 1. This result supports the conclusion that the observed increase in the SOT efficiency can be attributed to an enhancement of the spin-current absorption.

We have modelled our system as a trilayer of Pt/FM/Ru. When FM is thin enough, the spin current injected through the Pt/FM interface is not entirely absorbed in the FM layer, so that an additional torque is present at the FM/Ru interface. This additional torque, expressed through the *transmitted* mixing conductance, partially compensates the torque at Pt/FM interface. We obtain an expression of the damping torque [36,48],

$\tau = -\frac{j_c}{\sigma} 4\lambda_{Pt} \eta_{Pt} \theta_{Pt} \frac{1-\cosh^{-1} d_{Pt}/\lambda_{Pt}}{\tanh d_{Pt}/\lambda_{Pt}} \left[ \operatorname{Re} G_r^{\uparrow\downarrow,Pt} - \eta_{Ru} \operatorname{Re} G_t^{\uparrow\downarrow} \right]$, where $j_c$ is the current density, σ is the conductivity of the multilayer. $d_N$, $\lambda_N$ and $\sigma_N$ are the thickness, spin relaxation length, and conductivity of layer N, respectively, and $\eta_N = \dfrac{\sigma_N \tanh d_N / \lambda_N}{\sigma_N \tanh d_N / \lambda_N + 4\lambda_N \operatorname{Re} G_r^{\uparrow\downarrow,N}}$ is the transparency at N/FM interface. The torques at the Pt/FM and FM/Ru interfaces are expressed through the *reflected* and *transmitted* mixing conductance, $G_r^{\uparrow\downarrow,N}$ and $G_t^{\uparrow\downarrow}$, respectively. This formula shows a clear competition between the two interfacial torques. When the thickness of the FM is of the order or smaller than the spin dephasing length, the transmitted mixing conductance is finite and the torque at the top interface compensates the torque arising from Pt. In the presence of interfacial transverse spin current absorption, the spin current at the top interface vanishes ($G_t^{\uparrow\downarrow} \to 0$) and the total torque is solely driven by the torque from Pt $\sim G_r^{\uparrow\downarrow}$. While this theory qualitatively explains the experimental results, it is not sufficient to quantitatively describe the observed dependences on materials. In fact, parameters such as the saturation magnetization, interfacial structure, current distribution in the stack, etc. strongly depend on the FM thickness, which hinders quantitative interpretation.

In conclusion, this study presents a novel approach to largely manipulate the SOTs using a Ru capping layer. It is found that a thin Ru capping layer can greatly boost the spin current absorption into the ultrathin FM layer. In addition to the previous approaches to tune SOTs by varying the HM or FM layers, our results highlight that the modulating the torques from bottom and top interfaces of FM can substantially enhance the efficiency of SOTs. Moreover, together with the exchange coupling torque [35], the largely enhanced SOT in Pt/FM/Ru can be readily implemented to drive efficient magnetization switching and domain wall movement in Pt/SAF structure for various practical spintronic applications such as a MRAM and race-track domain wall memory.

This research is supported by the National Research Foundation (NRF), Prime Minister's Office, Singapore, under its Competitive Research Programme (CRP award no. NRFCRP12-2013-01). H.Y. is a member of the Singapore Spintronics Consortium (SG-SPIN). A.M. acknowledges support from King Abdullah University of Science and Technology (KAUST).

Fig. 1. (a) Film structure of the multilayers. (b) $R_H$ (offset for clarity) as a function of in-plane pulsed currents with different Ru thicknesses ($t_{Ru}$). $t_{Ru}$ is indicated for each curve. (c) Switching current density, $J_S$ (defined as the average absolute value of switching current densities between the high to low $R_H$ and vice-versa) as a function of $t_{Ru}$. (d) The SOT switching efficiency ($\eta$) as a function of $t_{Ru}$. (e) Longitudinal ($H_L$) and transverse ($H_T$) effective fields vs. $t_{Ru}$.

Fig. 2. $R_H$ (offset for clarity) as a function of pulsed currents with different FM thicknesses ($t_{FM}$) with a 0.6 nm Ru (a), and 2 nm MgO (b) capping layer. $t_{FM}$ is indicated for each curve. (c) The SOT switching efficiency ($\eta$). (d) Longitudinal ($H_L$) effective field. (e) Transverse ($H_T$) effective field as a function of $t_{FM}$. The solid (open) symbols in (c-e) represent the data with Ru (MgO) capping layer.

Fig. 3. (a) In a Pt/FM/MgO structure, the polarized spins from the Pt layer are partially reflected at the FM/MgO interfaces and compensate the torque at the Pt/FM interface. (b) In a Pt/FM/Ru structure, with strong spin relaxation near the FM/Ru interface, the absorption of spin currents from the Pt layer into FM is greatly enhanced. The white arrows denote the spin currents.

Fig. 4. (a) Schematic of the field modulated FMR setup. (b) FMR spectra with $t_{Ru}$ = 0 and 0.6 nm. (c) Full-width at half maximum (FWHM) at different frequencies to extract the damping constant ($\alpha$) for the samples with $t_{Ru}$ = 0, 0.6, and 1.4 nm. The inset shows the data from the reference sample. (d) Enhanced damping ($\Delta\alpha$) vs. $t_{Ru}$.

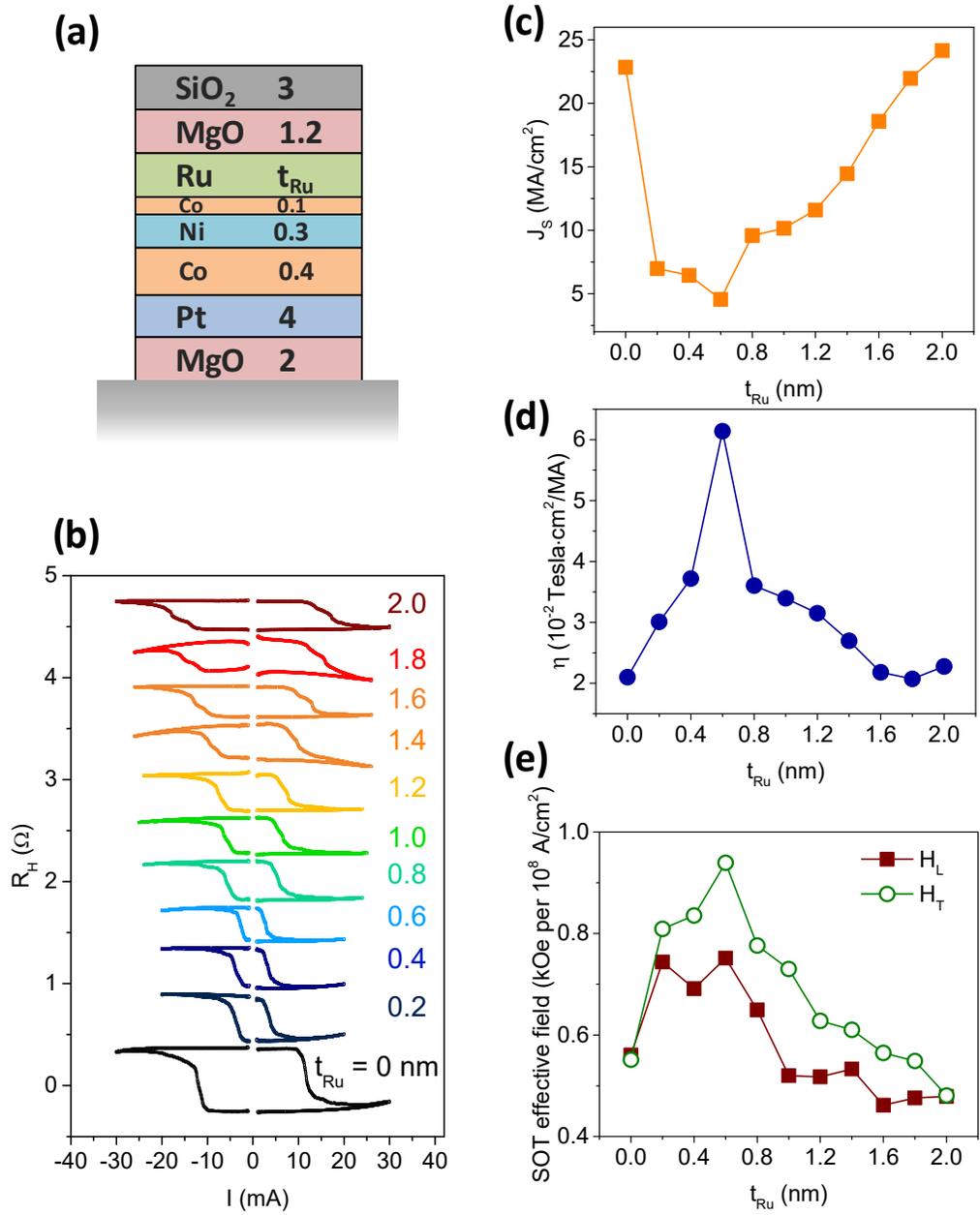

Figure 1

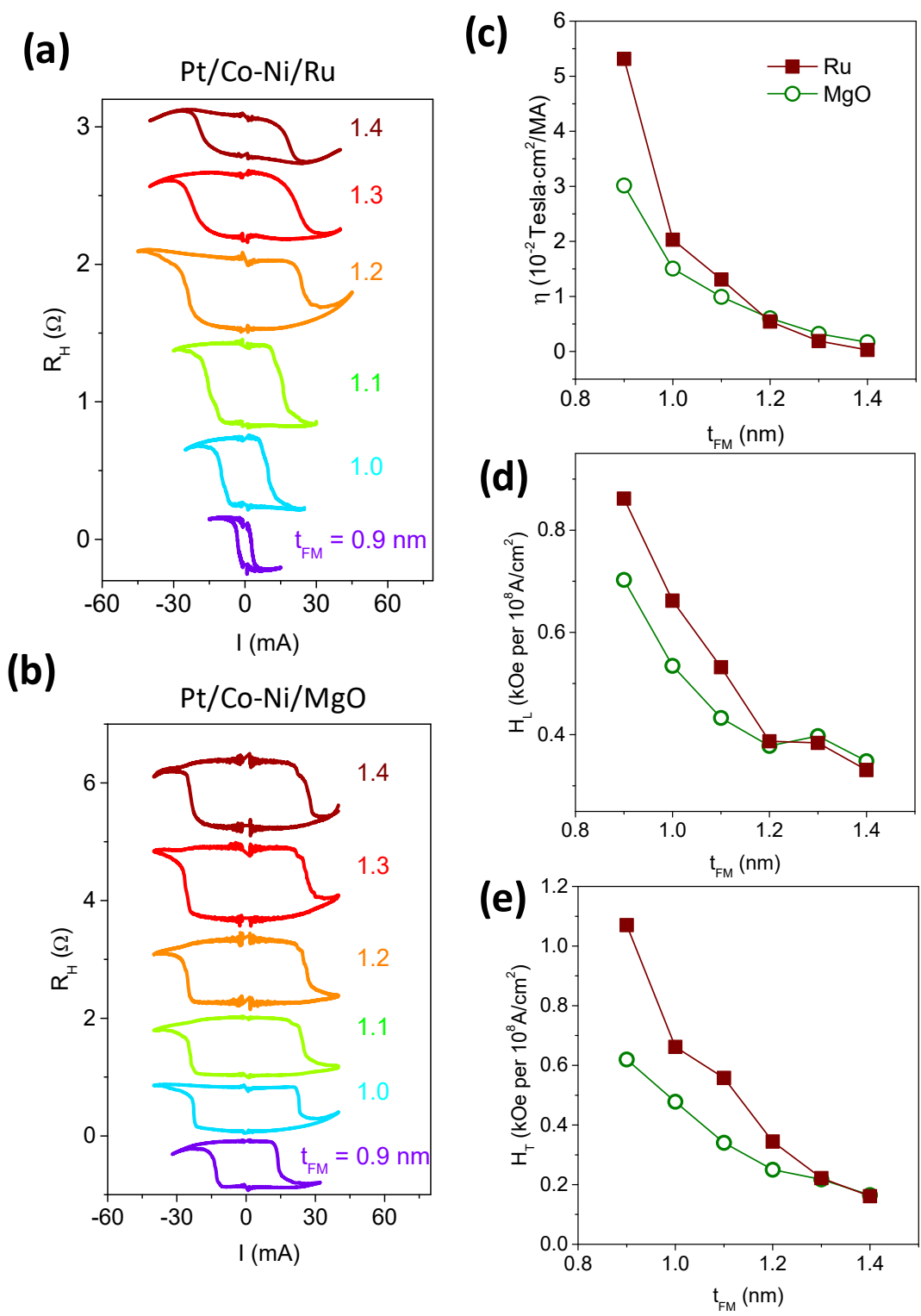

Figure 2

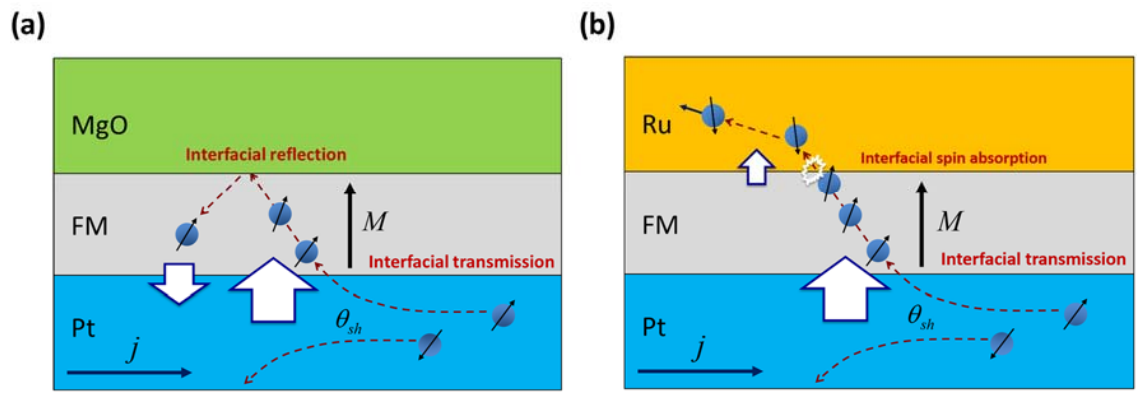

Figure 3

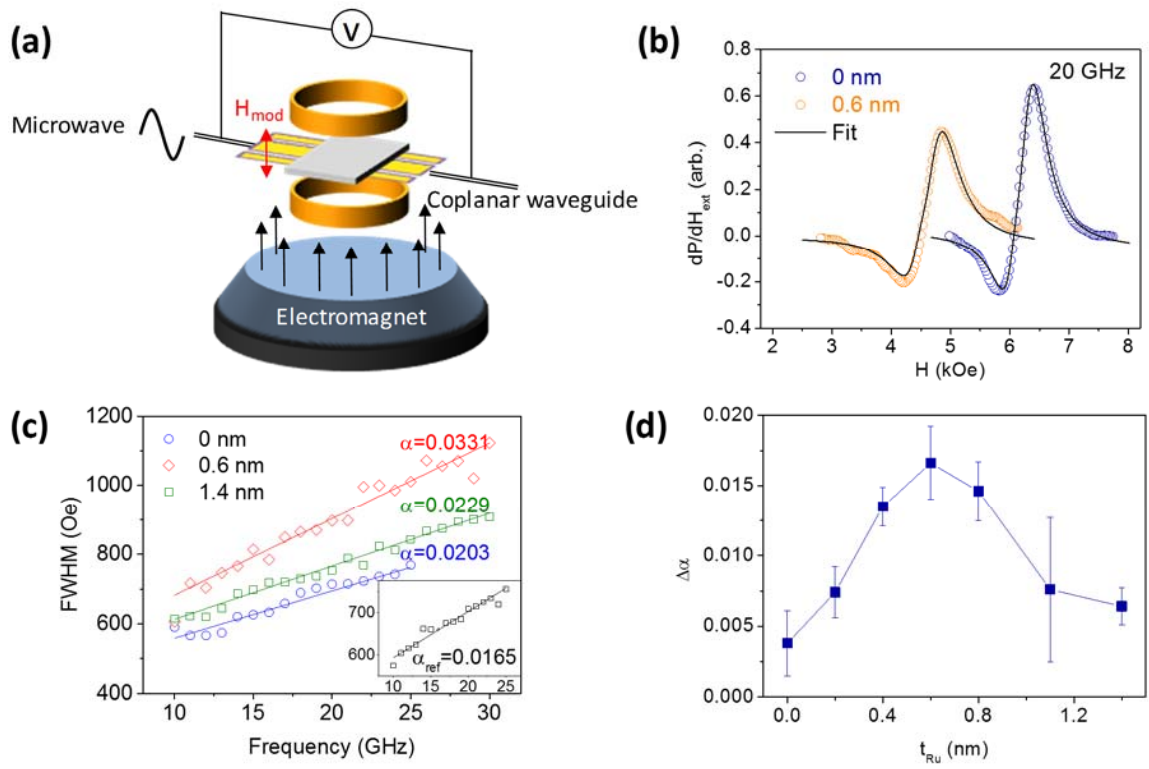

Figure 4